\documentclass[%
 reprint,
 amsmath,amssymb,
 aps,
prb,
 longbibliography
]{revtex4-2}

\usepackage[dvipdfmx]{graphicx}
\usepackage{dcolumn}
\usepackage{bm}
\usepackage{color}
\usepackage[usenames]{xcolor}
\usepackage{siunitx}
\usepackage{makecell}
\usepackage{hyperref}

\usepackage{ulem}
\usepackage[usenames]{xcolor}
\usepackage{epstopdf}

\begin{document}

\title{Metallic ruthenium ilmenites: first-principles study of MgRuO$_3$ and CdRuO$_3$}

\author{Seong-Hoon Jang}
\affiliation{
	National Institute for Materials Science, Tsukuba 305-0044, Japan
}
\author{Yukitoshi Motome}
\affiliation{
	Department of Applied Physics, University of Tokyo, Tokyo 113-8656, Japan
}

\date{\today}

\begin{abstract}
Ilmenites $AB$O$_3$ provide a platform for electron correlation and magnetism on alternatively stacked honeycomb layers of edge-sharing $A$O$_6$ or $B$O$_6$ octahedra. 
When $A$ and $B$ are $3d$ transition metals, strong electron correlation makes the systems Mott insulators showing various magnetic properties, while when $B$ is Ir with $5d$ electrons, competition between electron correlation and spin-orbit coupling realizes a spin-orbital coupled Mott insulator as a potential candidate for quantum spin liquids. 
Here we theoretically investigate intermediate $4d$ ilmenites, $A$RuO$_3$ with $A$=Mg and Cd, which were recently synthesized and shown to be metallic, unlike the $3d$ and $5d$ cases. 
By using first-principles calculations, we optimize the lattice structures and obtain the electronic band structures. 
We show that MgRuO$_3$ exhibits strong dimerization on RuO$_6$ honeycomb layers, leading to the formation of bonding and anti-bonding bands for one of three $t_{2g}$ orbitals; the lattice symmetry is lowered from $R\bar{3}$ to $P\bar{1}$, and the Fermi surfaces are composed of the other two $t_{2g}$ orbitals. 
In contrast, we find that CdRuO$_3$ has a lattice structure close to $R\bar{3}$, and all three $t_{2g}$ orbitals contribute almost equally to the Fermi surfaces. 
Comparison of our results with other Ru honeycomb materials such as Li$_2$RuO$_3$ indicates that the metallic ruthenium ilmenites stand on a subtle balance among electron correlation, spin-orbit coupling, and electron-phonon coupling. 
\end{abstract}


\maketitle 

\section{Introduction}
\label{sec:introduction}

\begin{figure*}[th!]
\includegraphics[width=2.0\columnwidth]{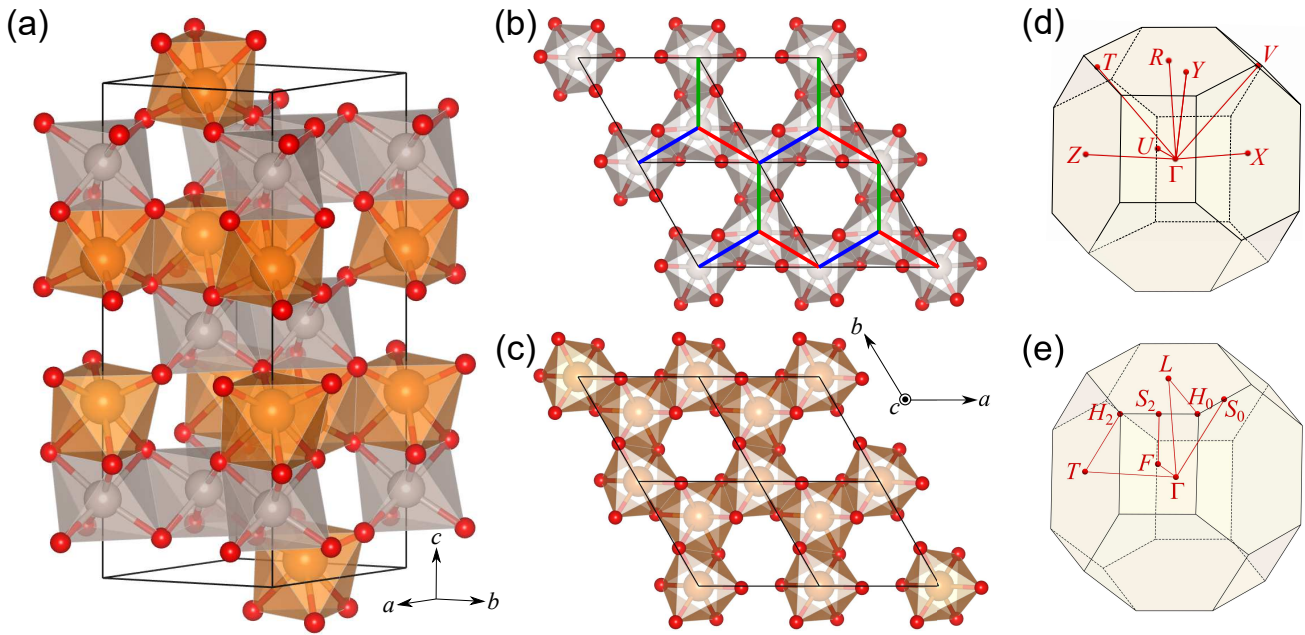} 
\caption{\label{fig:ilmenite}
(a) Lattice structure of ilmenite $A$RuO$_3$ ($A$= Mg and Cd). 
The gray and orange octahedra represent RuO$_6$ and $A$O$_6$, respectively.
(b) and (c) Top views of honeycomb networks of edge-sharing RuO$_6$ and $A$O$_6$ octahedra, respectively. 
The black rhombi denote the conventional unit cells, and the thick blue, red, and green lines in (b) represent the three different types of Ru-Ru bonds, dubbed $x$, $y$, and $z$ bonds; see Fig.~\ref{fig:RuO6_network}.
(d) and (e) First Brillouin zones for triclinic $P\overline{1}$ and rhombohedral $R\overline{3}$ symmetry, respectively.
The red lines represent the symmetric lines, along which the band structures are plotted in Figs.~\ref{fig:MgRuO3_fullyrelaxed}, \ref{fig:MgRuO3_others}, and \ref{fig:CdRuO3_fullyrelaxed}. 
}
\end{figure*}

Ilmenite-type crystal structure, whose name originally came from the titanium iron oxide mineral FeTiO$_3$, is widely found in materials with chemical formula $AB$O$_3$. 
The chemical formula is common to perovskite, and which structure is stable is empirically predicted by the so-called tolerance factor given by the ionic radii of $A$ and $B$ ions; 
ilmenite-type structure is in general stable for small tolerance factor with comparable ionic sizes of $A$ and $B$~\cite{Goldschmidt1926, Liu2009}.  
While perovskite has a three-dimensional cubic or rhombohedral network with corner-sharing $B$O$_6$ octahedra, ilmenite shows a two-dimensional layered structure of honeycomb network with edge-sharing octahedra. 
The layered honeycomb structure is commonly seen in corundum and LiNbO$_3$-type structures, but the arrangement of $A$O$_6$ and $B$O$_6$ as well as the lattice symmetry is different among them; 
ilmenite has a structure with alternating stacking of $A$O$_6$ and $B$O$_6$ honeycomb layers, as shown in Fig.~\ref{fig:ilmenite}. 

Among many ilmenite materials, $A$TiO$_3$ with transition metal ions for $A$ have been intensively studied for the rich magnetic properties from the $A$ magnetic ions. 
For instance, MnTiO$_3$ shows a N\'eel-type antiferromagnetic order with out-of-plane magnetic anisotropy, while NiTiO$_3$ and CoTiO$_3$ show a layered-type antiferromagnetic order with in-plane magnetic anisotropy~\cite{Shirane1959, Newnham:a04103, PhysRev.164.765, AKIMITSU197087, doi:10.1143/JPSJ.42.462, WATANABE1980549, PhysRevB.103.134438}.
In addition, chemically mixed compounds like Fe$_x$Mn$_{1-x}$TiO$_3$ and Ni$_x$Mn$_{1-x}$TiO$_3$ have also been studied for their spin glass behaviors due to the randomness and magnetic frustration~\cite{PhysRevLett.57.483, PhysRevLett.59.2364, doi:10.1143/JPSJ.58.1416, doi:10.1143/JPSJ.62.2575, doi:10.1143/JPSJ.63.3145, PhysRevLett.108.057203}. 

Recently, a series of ilmenite materials with $B$=Ir has been synthesized for $A$=Mg, Zn, and Cd~\cite{PhysRevMaterials.2.054411, PhysRevMaterials.4.044401}. 
In these iridium ilmenites $A$IrO$_3$, Ir ions nominally take the valence of Ir$^{4+}$, each of which has five electrons in the $5d$ orbitals on average. 
Since the relativistic spin-orbit coupling is expected to be substantial in the $5d$ electrons, these materials have attracted attention as good candidates for realizing the so-called spin-orbital coupled Mott insulator with Kitaev-type bond-dependent interactions, like Na$_2$IrO$_3$ and Li$_2$IrO$_3$~\cite{PhysRevLett.105.027204, SI2010, SI2012, PhysRevB.88.035107, Katukuri_2014, PhysRevLett.113.107201, WI2016, Freund2016, WI2017, TA2019, doi:10.7566/JPSJ.89.012002}. 
This picture was theoretically supported by first-principles calculations and perturbation calculations of the magnetic interactions for MgIrO$_3$ and ZnIrO$_3$~\cite{PhysRevMaterials.5.104409}. 

All of the above ilmenites with transition metal ions for $B$ are magnetic insulators. 
For $A$TiO$_3$, there are also materials with nonmagnetic ions $A$=Mg, Zn, and Cd, but they are band insulators or semiconductors, since Ti$^{4+}$ ions are nominally in the $3d^0$ state~\cite{doi:10.1143/JPSJ.13.1110, doi:10.1143/JPSJ.13.1298, KENNEDY20112987, doi:10.1021/ja408931v, Akaogi2015, Bahloul2017, doi:10.1021/acsaem.9b00815, doi:10.1063/5.0078021}. 
Very recently, however, there were experimental reports of probably the first metallic ilmenite MgRuO$_3$, where each Ru$^{4+}$ ion has four $4d$ electrons~\cite{Katori2021}.  
The material has metallic conductivity for all temperatures, but exhibits a phase transition at $T^* \sim 360$~K. 
The magnetic susceptibility is almost temperature independent above $T^*$, whereas it is reduced below $T^*$, suggesting a change of the Fermi surfaces at the phase transition. 
While the detailed properties as well as the origin of the phase transition have not yet been studied experimentally, it is interesting to clarify what type of metallic state is realized in this material. 
In particular, as the behaviors of the electrical conductivity and the magnetic susceptibility are qualitatively similar to the so-called spin-orbital coupled metals like Cd$_2$Re$_2$O$_7$ and PbRe$_2$O$_6$, where the phase transitions are accompanied by lattice deformations~\cite{doi:10.7566/JPSJ.87.024702, TAJIMA2020121359}, it would be intriguing to study the interplay between electronic properties and lattice structures in this unique metallic ilmenite with $4d$ electrons.

In this paper, we study the structural and electronic properties of the metallic ilmenite MgRuO$_3$ by using first-principles calculations. 
For comparison, we also study CdRuO$_3$ with a larger ionic size of the $A$-site cation. 
By the structural optimization, we show that MgRuO$_3$ exhibits strong dimerization in the Ru honeycomb plane, while CdRuO$_3$ almost preserves $C_3$ rotational symmetry. 
Accordingly, we find that the $4d$ $t_{2g}$ orbitals contribute differently to the electronic band structure between the two compounds: 
In MgRuO$_3$, the dimerization results in the formation of bonding and anti-bonding orbitals in one of the three $t_{2g}$ orbitals, and the Fermi surfaces are composed of the other two, whereas all the three $t_{2g}$ bands contribute almost equally to the Fermi surfaces in CdRuO$_3$. 
We discuss the dimerization in MgRuO$_3$ in comparison with Li$_2$RuO$_3$ which has similar honeycomb network of Ru ions~\cite{doi:10.1143/JPSJ.76.033705, doi:10.1143/JPSJ.78.094706}.

This paper is organized as follows. 
In Sec.~\ref{sec:Method}, we introduce the first-principles method that we use here, together with the details of the calculation conditions. 
We also describe the way of finding the Fermi surfaces from the first-principles results. 
In Sec.~\ref{sec:Results}, we show the results of lattice structures and electronic band structures for MgRuO$_3$ (Sec.~\ref{subsec:MgRuO3}) and CdRuO$_3$ (Sec.~\ref{subsec:CdRuO3}). 
Finally, Sec.~\ref{sec:Conclusion} is devoted to the summary and concluding remarks.

\section{Method}
\label{sec:Method}

We perform the first-principles calculations with the local density approximation including the nonrelativistic effect (LDA+SOC) by using \texttt{Quantum ESPRESSO}~\cite{GI2017}.
We first fully optimize the lattice structures (site positions and lattice translation vectors at once), starting from the experimental structure of MgIrO$_3$~\cite{PhysRevMaterials.2.054411} with replacement of Ir by Ru and Mg also by Cd for $A$=Cd. 
In the optimization, the minimum ionic displacement is set to 0.001~\si{\angstrom} in the Broyden-Fletcher-Goldfarb-Shanno iteration scheme~\cite{BR1970}.
In addition, we also prepare the symmetrized structures from the fully-optimized ones by modifying the lattice translation vectors of the conventional unit cell to the rhombohedral case and the site positions to the $R\overline{3}$ Wyckoff positions.
To examine the effect of lattice distortions, for the symmetrized structures, we modulate the length of one of the three types of bonds on each honeycomb plane by hand; in this procedure, the rhombohedral lattice translation vectors and the two Ru sites are fixed in the modulated structure, while the site positions of the $A$ and O ions are relaxed with the same minimum ionic displacement $0.001$~\si{\angstrom}.
For all the lattice structures, we calculate the electronic band structures and construct the maximally-localized Wannier functions (MLWFs) of Ru $4d$ $t_{2g}$ and O $2p$ orbitals by using \texttt{WANNIER90}~\cite{MO2014}.
Using the MLWFs, we calculate the projected density of states (DOS) and the Fermi surfaces for each orbital.
 
Throughout the calculations, we adopt the primitive unit cell which includes two $A$, two Ru, and six O ions, and the fully-relativistic and non-relativistic projector-augmented-wave-method Perdew-Zunger type pseudopotentials for the $A$ and O ions, respectively~\cite{PE1981, BL1994, DalCorso2014}.  
We take $4\times4\times4$ and $8\times8\times8$ Monkhorst-Pack $\mathbf{k}$-grids in the self-consistent field calculations for the structure optimizations and the non self-consistent field calculations for obtaining the electronic band structures and MLWFs, respectively~\cite{MO1976}; the convergence threshold for the self-consistent field calculations and the kinetic energy cutoff are set to $10^{-10}$~Ry and $200$~Ry, respectively.
We assume paramagnetic states for both $A$=Mg and Cd.

In the calculations of the Fermi surfaces, we construct the tight-binding Hamiltonians from the MLWFs, and obtain the eigenvalues and eigenvectors on $20\times20\times20$ $\mathbf{k}$-grids in a parallelepiped in momentum space spanned by the three reciprocal vectors corresponding to the lattice translations vectors of a primitive unit cell. 
Tracing the adiabatically connected eigenvectors, we identify the Fermi surfaces from the sign changes of their eigenvalues (the Fermi level is set to zero); with the aid of linear interpolation, the Fermi surfaces are drawn within the energy precision of $\pm10^{-8}$~eV. 
We also plot a two-dimensional cut of the Fermi surfaces on the plane including the $\Gamma$ point and perpendicular to the $\Gamma$-$Z$ ($\Gamma$-$T$) line for the $P\overline{1}$ ($R\overline{3}$) symmetry [see Figs.~\ref{fig:ilmenite}(d) and \ref{fig:ilmenite}(e)].
The procedure of the calculations is common to the above, but we take $50\times50$ $\mathbf{k}$-grids on a parallelogram with the minimum size including the relevant cross-section of the first Brillouin zone.

We also perform two additional calculations for MgRuO$_3$. 
One is the LDA+SOC calculations for $2\times 2\times 1$ primitive unit cell to examine the dimerization pattern in the honeycomb plane. 
The other is the calculations by the generalized gradient approximation (GGA+SOC) for the primitive unit cell to examine the correlation effect beyond LDA. 
The calculation setup is common to the LDA+SOC case, except for fully-relativistic and non-relativistic projector-augmented-wave-method Perdew–Burke–Ernzerhof type pseudopotentials for the Mg (Ru) and O ions, respectively~\cite{PhysRevLett.77.3865, DalCorso2014}, and the convergence threshold for the self-consistent field calculations set to $10^{-5}$~Ry. 
In the following sections, we show the results of the LDA+SOC calculations for the primitive unit cell unless otherwise noted.

\section{Results}
\label{sec:Results}

\subsection{MgRuO$_3$}
\label{subsec:MgRuO3}

\subsubsection{Lattice structure}
\label{subsubsec:MgRuO3LS}

\begin{figure}
\includegraphics[width=1.0\columnwidth]{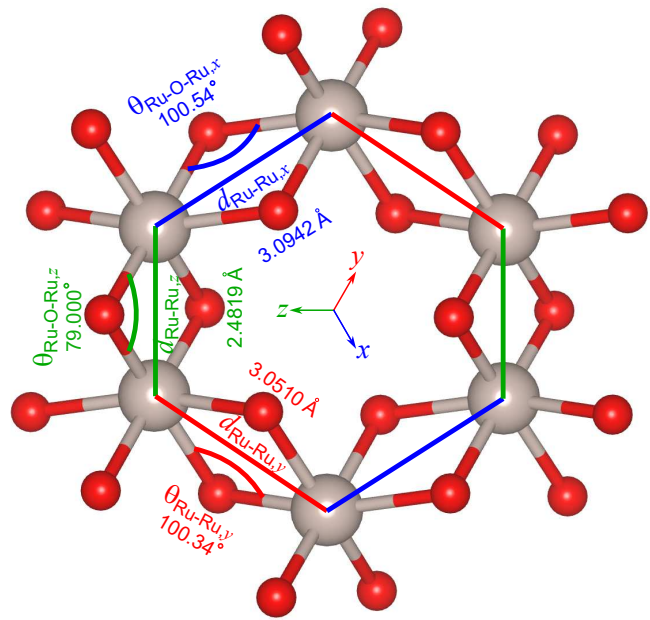} 
\caption{\label{fig:RuO6_network}
Fully-optimized lattice structure of MgRuO$_3$ displayed on a hexagon of RuO$_6$ octahedra. 
The symmetry is triclinic $P\overline{1}$.
The gray and red spheres represent Ru and O ions, respectively.
The hexagonal structure is composed of the three different Ru-Ru bonds, where the blue, red, and green lines represent the $x$, $y$, and $z$ bonds, respectively. 
We take the shortest Ru-Ru bond as the $z$ bond. 
The values of the Ru-Ru bond lengths $d_{\textrm{Ru-Ru,}\mu}$ and the Ru-O-Ru bond angles $\theta_{\textrm{Ru-O-Ru,}\mu}$ are shown. 
The $xyz$ axes are also shown, which are used for the definitions of $d_{yz}$, $d_{zx}$, and $d_{xy}$ orbitals later.
}
\end{figure} 

The full optimization of the lattice structure for MgRuO$_3$ converges with the triclinic $P\overline{1}$ symmetry.
The results are presented in Fig.~\ref{fig:RuO6_network} for a hexagon of RuO$_6$ octahedra. 
We find a considerable deviation from the $R\overline{3}$ symmetry in which the three types of Ru-Ru bonds on each honeycomb plane are all equivalent; one bond becomes about $20$~\% shorter than the other two. 
We label the shorter one as the $z$ bond and the other two as the $x$ and $y$ bonds. 
Accordingly, the Ru-O-Ru bond angle for the $z$ bond becomes smaller than $90\si{\degree}$, while those for the $x$ and $y$ bonds become larger, as shown in Fig.~\ref{fig:RuO6_network}. 
Despite this strong dimerization, each bond preserves the inversion symmetry with respect to the bond center.
We confirm that the GGA+SOC calculation reproduces the similar lattice distortion, while the dimerization is weaker; see Appendix~\ref{appA}.

The dimerization reminds us of similar but different type of dimerization in Li$_2$RuO$_3$~\cite{doi:10.1143/JPSJ.76.033705}, which has also layered honeycomb structures of edge-sharing RuO$_6$ octahedra with the same nominal valence of Ru$^{4+}$ as MgRuO$_3$. 
The dimerized pattern in Li$_2$RuO$_3$ is not compatible with the primitive unit cell, and has $2\times 2\times 1$ periodicity. 
We performed the LDA+SOC calculation for $2\times 2\times 1$ primitive unit cell and confirmed that MgRuO$_3$ does not prefer to the pattern in Li$_2$RuO$_3$ but a simple repetition of Fig.~\ref{fig:RuO6_network} with $1\times 1\times 1$ periodicity.

\begin{figure}
\includegraphics[width=1.0\columnwidth]{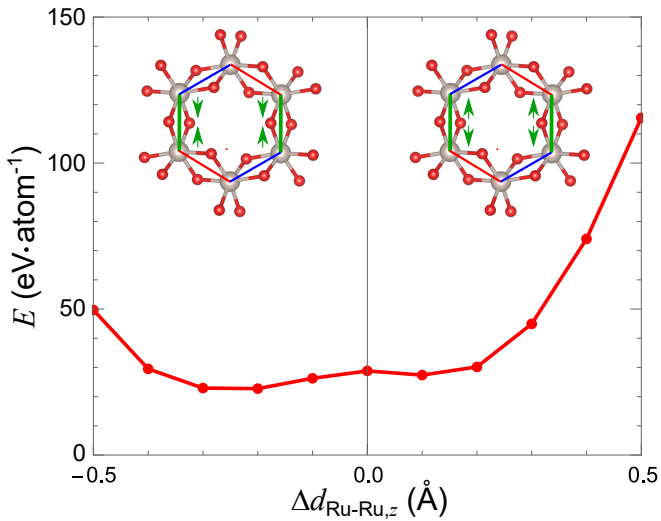} 
\caption{\label{fig:MgRuO3_EDFT}
Energy obtained by the LDA+SOC calculations while varying the Ru-Ru bond length on the $z$ bond for MgRuO$_3$.
The energy is measured from that of the fully-optimized lattice structure with $P\overline{1}$ symmetry, and the bond length is measured from $d_{\textrm{Ru-Ru,}z}=2.8973$~\si{\angstrom} for the symmetrized $R\overline{3}$ structure. 
The insets show the schematics of the contraction (left) and elongation (right) of the $z$ bond.
}
\end{figure} 

To further examine the stability of the dimerized structure, we evaluate the energy while changing the Ru-Ru bond length on the $z$ bond by $\Delta d_{\textrm{Ru-Ru,}z}$. 
This is done with respect to the symmetrized structure with the rhombohedral $R\overline{3}$ symmetry with $d_{\textrm{Ru-Ru,}z}=2.8973$~\si{\angstrom}, which is obtained by the symmetrization for the fully-optimized structure, following the procedure described in Sec.~\ref{sec:Method}. 
The result is shown in Fig.~\ref{fig:MgRuO3_EDFT}.
Note that the symmetrized $R\overline{3}$ structure with $\Delta d_{\textrm{Ru-Ru,}z} = 0$ gives higher energy than the fully-optimized one by $\simeq 30$~meV$\cdot$atom$^{-1}$, which is still moderately unstable as being far less than 100~meV$\cdot$atom$^{-1}$~\cite{PhysRevMaterials.2.043802}.
Here, negative and positive $\Delta d_{\textrm{Ru-Ru,}z}$ correspond to the contraction and elongation of the $z$ bond, respectively. 
Both modulations reduce the $R\overline{3}$ symmetry to the $P\overline{1}$ symmetry; the contraction leads to dimerization similar to the fully-optimized structure, while the elongation leads to a quasi-one-dimensional zigzag structure composed of the shorter $x$ and $y$ bonds (see the insets of Fig.~\ref{fig:MgRuO3_EDFT}).
We find two meta-stable local minima with nonzero lattice distortions, $\Delta d_{\textrm{Ru-Ru,}z} \simeq -0.25$~\si{\angstrom} and $\simeq 0.1$~\si{\angstrom}, as shown in Fig.~\ref{fig:MgRuO3_EDFT}. 
The energy is slightly lower for the former dimerized case rather than the latter zigzag one. 
While the degree of the dimerization is smaller than the fully-optimized one in Fig.~\ref{fig:RuO6_network}, the result rationalizes the instability toward the dimerization found in the full structural optimization.

\subsubsection{Electronic band structure}
\label{subsubsec:MgRuO3EBS}

\begin{figure}
\includegraphics[width=1.0\columnwidth]{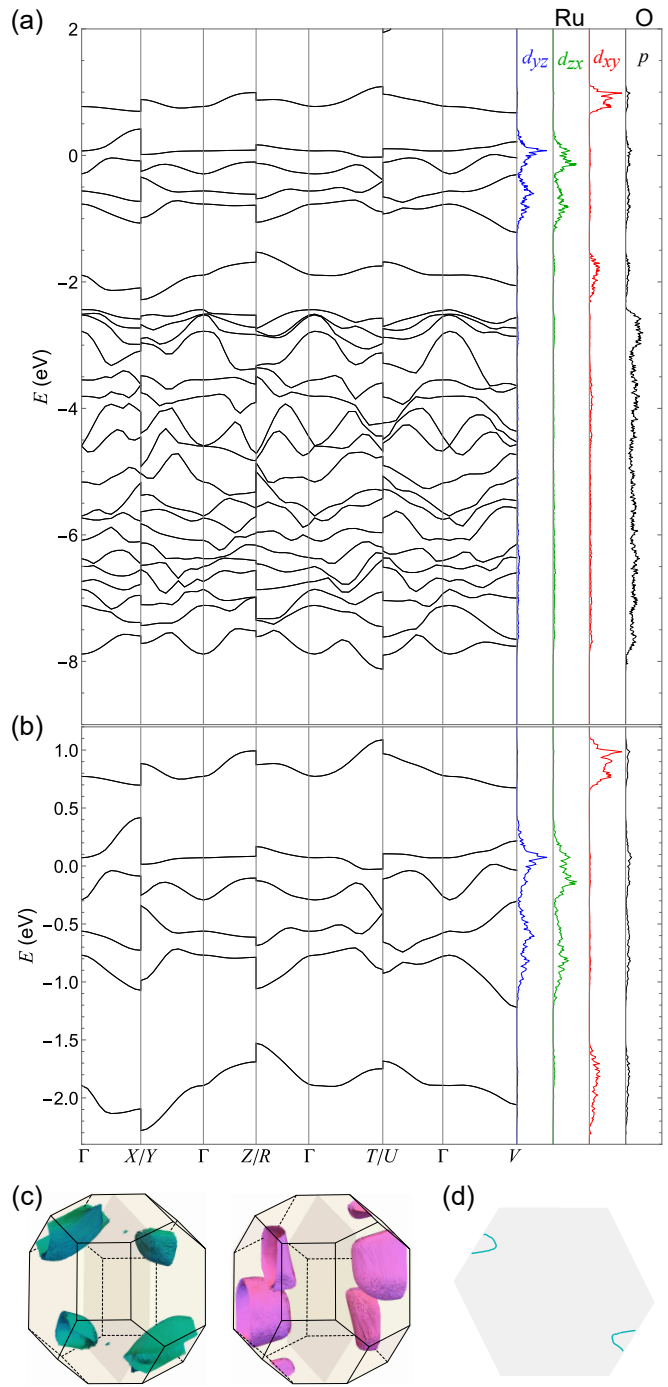} 
\caption{\label{fig:MgRuO3_fullyrelaxed}
(a) and (b) Electronic band structure of MgRuO$_3$ obtained by the LDA+SOC calculation for the fully-optimized structure with $P\overline{1}$ symmetry, displayed in the energy range of (a) [$-9.0$, $2.0$]~eV and (b) [$-2.4$, $1.2$]~eV. 
The right panels represent the projected DOS for the Ru $4d$ $t_{2g}$ and O $2p$ orbitals. 
The Fermi level is set to zero. 
The symmetric lines and the first Brillouin zones are given in Fig.~\ref{fig:ilmenite}(d).
(c) Fermi surfaces in the first Brillouin zone, for the second-highest (left) and third-highest (right) energy bands in (b).
(d) Two-dimensional cut of the Fermi surfaces on the gray cross-sections in (c).
}
\end{figure} 

\begin{figure*}[t!]
\includegraphics[width=2.0\columnwidth]{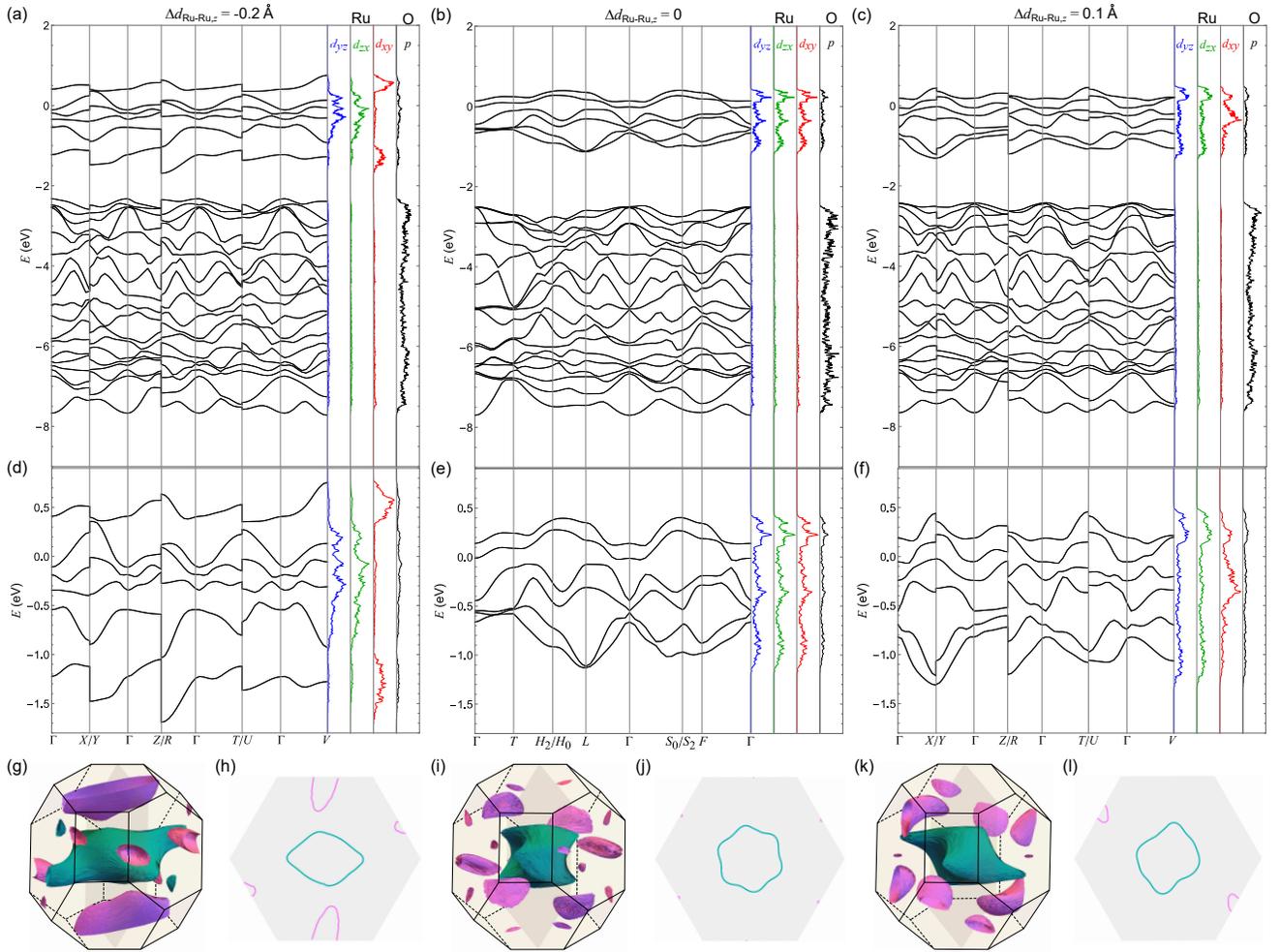} 
\caption{\label{fig:MgRuO3_others}
(a)--(f) Electronic band structures of MgRuO$_3$ with the modulation of the Ru-Ru bond length on the $z$ bond: (a) and (d) $\Delta d_{\textrm{Ru-Ru,}z}=-0.2$~\si{\angstrom} ($P\overline{1}$ symmetry), (b) and (e) $\Delta d_{\textrm{Ru-Ru,}z}=0$~\si{\angstrom} ($R\overline{3}$ symmetry), and (c) and (f) $\Delta d_{\textrm{Ru-Ru,}z}=0.1$~\si{\angstrom} ($P\overline{1}$ symmetry). 
See also Fig.~\ref{fig:MgRuO3_EDFT}. 
The symmetric lines and the first Brillouin zones are given in Figs.~\ref{fig:ilmenite}(d) and \ref{fig:ilmenite}(e) for the $P\overline{1}$ and the $R\overline{3}$ symmetries, respectively.
(g), (i), and (k) Fermi surfaces in the first Brillouin zone for each case. 
(h), (j), and (l) Two-dimensional cuts of the Fermi surfaces. 
The notations are common to those in Fig.~\ref{fig:MgRuO3_fullyrelaxed}.
}
\end{figure*} 

Figure~\ref{fig:MgRuO3_fullyrelaxed}(a) shows the electronic band structure for MgRuO$_3$ with the fully-optimized lattice structure with $P\overline{1}$ symmetry. 
As indicated in the projected DOS in the right panels, the energy bands near the Fermi level are dominated by the $4d$ $t_{2g}$ orbitals of Ru ions with weak hybridization with the O $2p$ orbitals. 
The enlarged figure in the energy range of $[-2.4, 1.2]$~eV is shown in Fig.~\ref{fig:MgRuO3_fullyrelaxed}(b). 
The most relevant bands traversing the Fermi level are dominated by the $d_{yz}$ and $d_{zx}$ orbitals.
The projected DOS for these two orbitals are similar to each other, reflecting the similar local structures of the $x$ and $y$ bonds in Fig.~\ref{fig:RuO6_network}.
In contrast, the $d_{xy}$ band is split by approximately $3$~eV due to the strong dimerization on the $z$ bond, forming the bonding and antibonding bands which locate well below and above the Fermi level, respectively. 
Thus, the results indicate that MgRuO$_3$ is metallic, where the Fermi surfaces are dominated by the $d_{yz}$ and $d_{zx}$ orbitals; there is no contribution from the $d_{xy}$ orbital which forms a band insulating state with fully-filled bonding band. 

We display the Fermi surfaces composed of the two orbitals in Fig.~\ref{fig:MgRuO3_fullyrelaxed}(c). 
Both of them form hole pockets across the zone boundary, satisfying the $P\overline{1}$ symmetry. 
In Fig.~\ref{fig:MgRuO3_fullyrelaxed}(d), we also present the two-dimensional cut of the Fermi surfaces on the gray cross-sections in Fig.~\ref{fig:MgRuO3_fullyrelaxed}(c), where only the second-highest energy band participates.

We also compute the band structures for the lattice structures obtained by changing $\Delta d_{\textrm{Ru-Ru,}z}$ in Fig.~\ref{fig:MgRuO3_EDFT}. 
Figure~\ref{fig:MgRuO3_others} displays the results for the dimerized case with $\Delta d_{\textrm{Ru-Ru,}z} = -0.2$~\si{\angstrom} [(a) and (d)], the perfect $R\overline{3}$ case with $\Delta d_{\textrm{Ru-Ru,}z} = 0$ [(b) and (e)], and the zigzag case with $\Delta d_{\textrm{Ru-Ru,}z} = 0.1$~\si{\angstrom} [(c) and (f)].
The results indicate systematic changes of the band structures with $\Delta d_{\textrm{Ru-Ru,}z}$. 
For $\Delta d_{\textrm{Ru-Ru,}z} = -0.2$~\si{\angstrom}, the $d_{yz}$ and $d_{zx}$ bands traverse the Fermi level, and the $d_{xy}$ band is split by approximately $2$~eV and located away from the Fermi level, similar to the fully-optimized case in Fig.~\ref{fig:MgRuO3_fullyrelaxed}.
Meanwhile, for $\Delta d_{\textrm{Ru-Ru,}z} = 0$, the projected DOS for the $d_{yz}$, $d_{zx}$, and $d_{xy}$ orbitals become identical, reflecting the $C_3$ symmetry in each honeycomb layer.
For $\Delta d_{\textrm{Ru-Ru,}z} = 0.1$~\si{\angstrom}, the elongated $d_{\textrm{Ru-Ru,}z}$ brings about asymmetry between $d_{xy}$ and ($d_{yz}, d_{zx}$), while all three bands contribute to the metallic conduction despite the formation of the zigzag chain structure.

We display the Fermi surfaces for each case in Figs.~\ref{fig:MgRuO3_others}(g), \ref{fig:MgRuO3_others}(i), and \ref{fig:MgRuO3_others}(k).
For $\Delta d_{\textrm{Ru-Ru,}z} = 0$, the Fermi surfaces respect the $C_3$ symmetry about the $\Gamma$-$T$ axis in Fig.~\ref{fig:ilmenite}(e), while for $\Delta d_{\textrm{Ru-Ru,}z} \neq 0$, they are distorted with broken $C_3$ symmetry.
We also display the two-dimensional cuts of the Fermi surfaces in Figs.~\ref{fig:MgRuO3_others}(h), \ref{fig:MgRuO3_others}(j), and \ref{fig:MgRuO3_others}(l).
The results clearly indicate that the Fermi surfaces respect the $C_3$ symmetry at $\Delta d_{\textrm{Ru-Ru,}z} = 0$, whereas they are distorted for the other cases.

\subsection{CdRuO$_3$}
\label{subsec:CdRuO3}

\subsubsection{Lattice structure}
\label{subsubsec:CdRuO3LS}

\begin{figure}
\includegraphics[width=1.0\columnwidth]{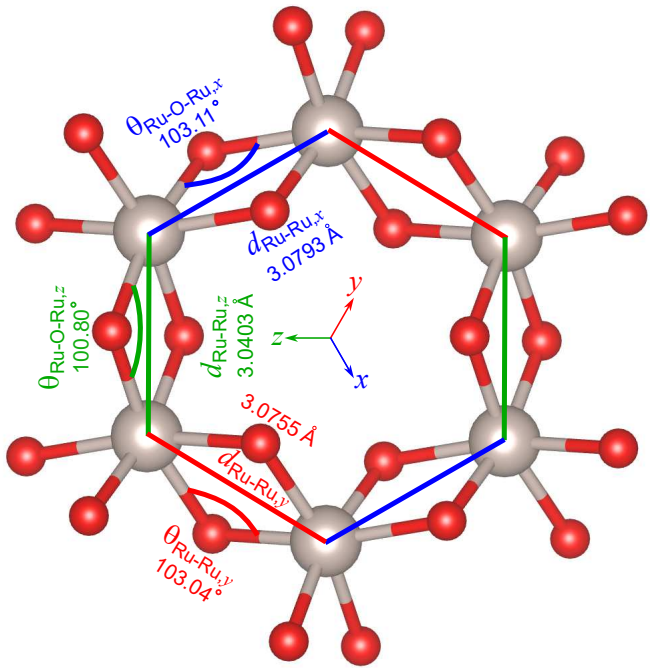} 
\caption{\label{fig:CdRuO6_network}
Fully-optimized lattice structure of CdRuO$_3$ displayed on a hexagon of RuO$_6$ octahedra. 
The symmetry is triclinic $P\overline{1}$, while the deviation from $R\overline{3}$ is very small. 
The notations are common to Fig.~\ref{fig:RuO6_network}. 
}
\end{figure} 

The full optimization of the lattice structure for CdRuO$_3$ converges with the triclinic $P\overline{1}$ symmetry, similar to MgRuO$_3$. 
The results are presented in Fig.~\ref{fig:CdRuO6_network}.
In this case, however, the deviation from $R\overline{3}$ symmetry is very small; the difference between the three types of bond lengths is about $1$~\%. 
In addition, each bond preserves the inversion symmetry with respect to the bond center, as in the case of MgRuO$_3$.

\begin{figure}[bh!]
\includegraphics[width=1.0\columnwidth]{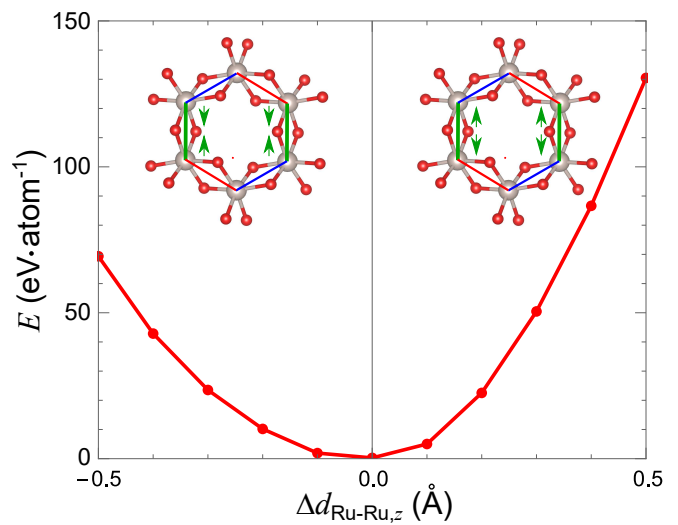} 
\caption{\label{fig:CdRuO3_EDFT}
Energy obtained by the LDA+SOC calculations while varying the Ru-Ru bond length on the $z$ bond for CdRuO$_3$. 
The energy is measured from that of the fully-optimized lattice structure with $P\overline{1}$ symmetry, and the bond length is measured from $d_{\textrm{Ru-Ru,}z}=3.0677$~\si{\angstrom} for the symmetrized $R\overline{3}$ structure.
The notations are common to Fig.~\ref{fig:MgRuO3_EDFT}.
}
\end{figure} 

Similar to the case of MgRuO$_3$ in Fig.~\ref{fig:MgRuO3_EDFT}, we evaluate the energy while changing $\Delta d_{\textrm{Ru-Ru,}z}$ with respect to the symmetrized $R\overline{3}$ structure with $d_{\textrm{Ru-Ru,}z}=3.0677$~\si{\angstrom} obtained from the fully-optimized structure.
The result is shown in Fig.~\ref{fig:CdRuO3_EDFT}.  
We find that the energy is minimized at the symmetric case with $\Delta d_{\textrm{Ru-Ru,}z}=0$ and there are no local minima with either contraction or elongation of the $z$ bonds, in contrast to MgRuO$_3$. 
Note that the minimum energy is very close to that for the fully-optimized one (the difference is $0.264$~meV$\cdot$atom$^{-1}$), as expected from the small deviation from the $R\overline{3}$ symmetry. 
 
\subsubsection{Electronic band structure}
\label{subsubsec:CdRuO3EBS}

\begin{figure}[th!]
\includegraphics[width=1.0\columnwidth]{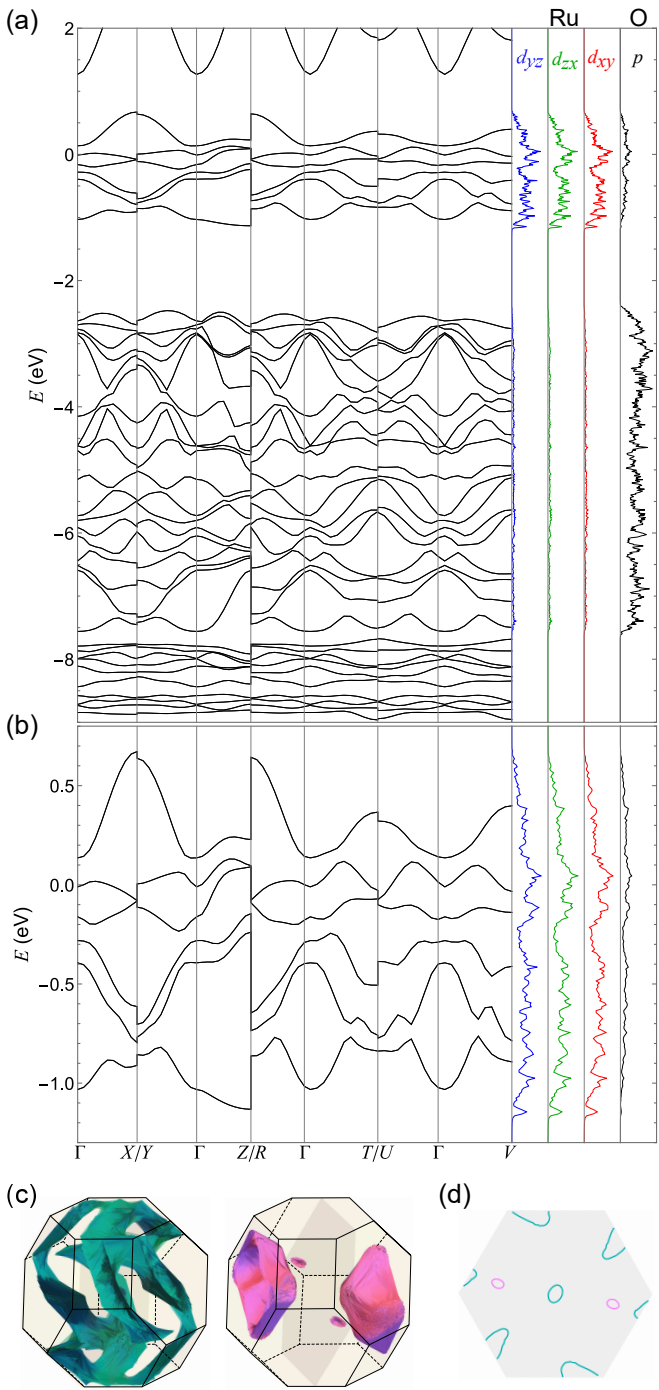} 
\caption{\label{fig:CdRuO3_fullyrelaxed}
(a) and (b) Electronic band structure of CdRuO$_3$ obtained by the LDA+SOC calculation for the fully-optimized structure with $P\overline{1}$ symmetry, displayed in the energy range of (a) [$-9.0$, $2.0$]~eV and (b) [$-1.3$, $0.8$]~eV.
(c) Fermi surfaces in the first Brillouin zone, for the second-highest (left) and third-highest (right) energy bands in (b). 
(d) Two-dimensional cuts of the Fermi surfaces.
The notations are common to Fig.~\ref{fig:MgRuO3_fullyrelaxed}.
}
\end{figure} 

Figure~\ref{fig:CdRuO3_fullyrelaxed}(a) shows the electronic band structure for CdRuO$_3$ with the fully-optimized lattice structure with $P\overline{1}$ symmetry. 
As indicated in the projected DOS in the right panels, the energy bands near the Fermi level are dominated by the $4d$ $t_{2g}$ orbitals of Ru ions with weak hybridization with the O $2p$ orbitals; as clearly shown in the enlarged figure in Fig.~\ref{fig:CdRuO3_fullyrelaxed}(b), the $d_{xy}$, $d_{yz}$, and $d_{zx}$ bands contribute almost equally, similar to the symmetrized case of MgRuO$_3$ shown in Figs.~\ref{fig:MgRuO3_others}(b) and \ref{fig:MgRuO3_others}(e). 
We display the Fermi surfaces of the two bands traversing the Fermi level in Fig.~\ref{fig:CdRuO3_fullyrelaxed}(c).
We also represent the two-dimensional cut of the Fermi surfaces in Fig.~\ref{fig:CdRuO3_fullyrelaxed}(d).
These results approximately respect the $C_3$ symmetry about the $\Gamma$-$T$ axis as expected from the small deviation from the $R\overline{3}$ symmetry. 
We do not show the results for the energy minimized state in Fig.~\ref{fig:CdRuO3_EDFT} since they would be very close to the results in Figs.~\ref{fig:CdRuO3_fullyrelaxed}(a) and \ref{fig:CdRuO3_fullyrelaxed}(b) because of the structural similarity.

\section{Concluding remarks}
\label{sec:Conclusion}

In summary, we have studied the electronic band structures of the metallic ruthenium ilmenites $A$RuO$_3$ with $A$=Mg and Cd, by using the first-principles calculations with structural optimization. 
For MgRuO$_3$, we found that the optimized lattice structure exhibits strong dimerization on one of the three bonds in the honeycomb network of edge-sharing MgO$_6$ octahedra. 
Because of the dimerization, one of three $4d$ $t_{2g}$ orbitals of Ru ions causes bonding and anti-bonding splitting, and the bonding band is fully filled; the Fermi surfaces are composed of the other two $t_{2g}$ orbitals. 
We also showed that MgRuO$_3$ has a metastable structure for the opposite lattice modification to the dimerization, namely, for the formation of coupled zigzag chains composed of the other two bonds, but the energy is higher than the dimerized one. 
Meanwhile, for CdRuO$_3$, we found that the optimized lattice structure is very close to the symmetric $R\overline{3}$ structure, resulting in almost equal contributions from the three $t_{2g}$ orbitals to the Fermi surfaces. 
The contrastive behaviors between $A$=Mg and Cd indicate that the ionic size of the $A$-site cation are relevant to the structural and electronic properties in the metallic ilmenites, as in the insulating magnets $A$IrO$_3$ with $A$=Mg, Zn, and Cd~\cite{PhysRevMaterials.4.044401}. 

In experiments, MgRuO$_3$ exhibits a phase transition at $T^* \sim 360$~K, as mentioned in Sec.~\ref{sec:introduction}. 
Although the detailed nature of this transition remains unexplored, it might be a structural transition from the high-temperature $R\overline{3}$ phase to the dimerized $P\overline{1}$ phase found in our results. 
If this is the case, our finding suggests that the DOS at the Fermi level is largely reduced as one of three $t_{2g}$ bands is gapped out by the formation of binding and anti-bonding bands associated with the dimerization. 
This accounts well for the reduction of the magnetic susceptibility at the phase transition on cooling. 
Detailed measurements on the lattice structures and the electronic states are needed to confirm this scenario.

Such a metal-metal transition reminds us of those in the spin-orbital coupled metals, such as Cd$_2$Re$_2$O$_7$~\cite{doi:10.7566/JPSJ.87.024702} and PbRe$_2$O$_6$~\cite{TAJIMA2020121359}. 
In these $5d$-electron systems, the phase transitions break spatial inversion symmetry and activate the antisymmetric spin-orbit coupling, which can induce odd-parity multipoles and anomalous transport phenomena~\cite{PhysRevLett.122.147602}. 
We note, however, that the dimerized $P\overline{1}$ state for MgRuO$_3$ in our results preserves the spatial inversion symmetry, and hence, such properties are not expected in this spin-orbital coupled metal.

Our results concluded that MgRuO$_3$ shows a different dimerized pattern from Li$_2$RuO$_3$.
Li$_2$RuO$_3$ exhibits a phase transition at $\sim 540$~K, where the magnetic susceptibility shows a sharp drop from almost temperature independent behavior, similar to MgRuO$_3$. 
At this transition, however, the electrical conductivity increases, suggesting a metal-insulator transition, in contrast to the metallic behavior both above and below the transition temperature in MgRuO$_3$. 
The subsequent first-principles calculation showed that a very small energy gap is formed by dimerization in Li$_2$RuO$_3$~\cite{doi:10.1143/JPSJ.78.094706}. 
The origin of dimerization was also studied from the strong correlation limit~
\cite{PhysRevLett.100.147203}.  
In contrast, our GGA+SOC calculation for MgRuO$_3$ predicted a metallic state, consistent with the experimental result~\cite{Katori2021}.
Given that the lattice parameters are similar between the metallic MgRuO$_3$ and the insulating Li$_2$RuO$_3$, the different electronic nature might be ascribed to subtle balance between electronic correlation, spin-orbit coupling, and electron-phonon coupling on the different underlying lattice symmetry in these $4d$ transition metal compounds. 
This interesting problem is left for future study. 

\begin{acknowledgments}
The authors thank fruitful discussions with Y. Haraguchi.
The crystal structures in Figs.~\ref{fig:ilmenite}(a), \ref{fig:ilmenite}(b), \ref{fig:ilmenite}(c), \ref{fig:RuO6_network}, and \ref{fig:CdRuO6_network} were visualized by \texttt{VESTA}~\cite{MO2011}. 
Reference~\cite{HINUMA2017140} was referred to for the Brillouin zone and the symmetric lines and points in Figs.~\ref{fig:ilmenite}(d) and \ref{fig:ilmenite}(e).
Parts of the numerical calculations have been done using the facilities of the Supercomputer Center, the Institute for Solid State Physics, the University of Tokyo. This work was supported by JST CREST (JP-MJCR18T2), and JSPS KAKENHI Grants No.~19H05825 and No.~20H00122.
S.-H. J thanks for the financial support from ``Program for Promoting Research on the Supercomputer Fugaku'' Grant No.~JPMXP1020200301.
\end{acknowledgments}

\appendix

\section{Optimized strucutre for MgRuO$_3$ with GGA+SOC}
\label{appA}

\begin{table}[bh!]
\caption{\label{tab:MgRuO3_GGA}
Ru-Ru bond lengths $d_{\textrm{Ru-Ru,}\mu}$ and Ru-O-Ru bond angles $\theta_{\textrm{Ru-O-Ru,}\mu}$ for $\mu$(=$x$, $y$, and $z$) bonds in the fully-optimized RuO$_6$ network of MgRuO$_3$ given by using GGA+SOC.
See Fig.~\ref{fig:RuO6_network} for the definitions of $d_{\textrm{Ru-Ru,}\mu}$ and $\theta_{\textrm{Ru-O-Ru,}\mu}$.
}
\begin{ruledtabular}
\begin{tabular}{ccccccc}
$\mu$ & $x$ & $y$ & $z$\\
\hline
$d_{\textrm{Ru-Ru,}\mu}$~(\si{\angstrom}) & $2.9410$ & $2.9247$ & $2.8694$\\
$\theta_{\textrm{Ru-O-Ru,}\mu}$~(\si{\degree}) & $93.070$ & $92.713$ & $90.815$\\
\end{tabular}
\end{ruledtabular}
\end{table}

In Table~\ref{tab:MgRuO3_GGA}, we present the lattice parameters of MgRuO$_3$ fully optimized with generalized gradient approximation including the relativistic effect (GGA+SOC). 
The lattice structure converges with the triclinic $P\overline{1}$ symmetry similar to the LDA+SOC case, while the degree of dimerization becomes weaker.


\bibliography{main}

\end{document}